\documentclass[journal]{IEEEtran}
\usepackage[T1]{fontenc}
\usepackage[utf8]{inputenc}
\usepackage{amsmath}

\usepackage{amssymb}
\usepackage{mathtools}
\usepackage[dvipsnames]{xcolor}
\usepackage{graphicx,balance}
\usepackage{capt-of}
\usepackage{textcomp}
\usepackage{url}
\usepackage{verbatim}
\usepackage{cite}
\usepackage{array}
\usepackage{multirow}
\usepackage{makecell}
\usepackage{enumitem}
\usepackage{xspace}
\usepackage{stfloats}
\usepackage{listings}  
\usepackage{ragged2e}
\usepackage{caption}
\usepackage[most]{tcolorbox}

\newtheorem{definition}{Definition}

\newcommand{\fu}[1]{\textcolor{red}{#1}}
\newcommand{\rk}[1]{\textcolor{blue}{#1}}

\newcommand{\tool}{{\sf RAINFALL}\xspace}
\newcommand{\toolv}{{\sf RAINFALL$^\star$}\xspace} 

\usepackage[hidelinks]{hyperref}

\begin{document}

\title{Ranking Function Generation with LLMs\\for Termination}
\author{Rongkang~Li, Jiaxin~Hu, Taolue~Chen, Yong~Li, Fu~Song, Zhilin~Wu and Qiusong Yang%
\thanks{Rongkang Li, Jiaxin Hu, Yong Li, Fu Song, Zhilin Wu and Qiusong Yang are with Key Laboratory of System Software (Chinese Academy of Sciences),  Institute of Software, Chinese Academy of Sciences, and University of Chinese Academy of Sciences, Beijing, China. Rongkang Li, Jiaxin Hu, and Fu Song are also with  Nanjing Institute of Software Technology, Nanjing, China (e-mail: lirongkang25@mails.ucas.ac.cn; hujiaxin26@mails.ucas.ac.cn; liyong@ios.ac.cn; songfu@ios.ac.cn; wuzl@ios.ac.cn; yangqs@ios.ac.cn; ORCID: 0009-0005-9954-2612, 0009-0002-5089-9440, 0000-0002-7301-9234, 0000-0002-0581-2679, 0000-0003-0899-628X, 0000-0001-8099-2035).}%
\thanks{Taolue Chen is with School of Computing and Mathematical Sciences, Birkbeck, University of London, London, U.K. (e-mail: t.chen@bbk.ac.uk; ORCID: 0000-0002-5993-1665).}%
\thanks{Corresponding authors: Yong Li and Fu Song.}%
\thanks{Manuscript received XX XXXX 20XX; revised XX XXXX 20XX; accepted XX XXXX 20XX.}}

\markboth{IEEE Transactions on Software Engineering,~Vol.~XX, No.~X, XXXXX~2026}%
{Li \MakeLowercase{\textit{et al.}}: Ranking Function Generation with LLMs for Termination}

\maketitle

\begin{abstract} 
Ranking functions are a fundamental technique for proving loop termination, but synthesizing them remains challenging and often requires substantial expertise. Despite considerable progress in their automated synthesis, existing approaches remain limited in their effectiveness. We propose \tool, a neuro-symbolic framework that tightly integrates large language models (LLMs) with symbolic verification. To our best knowledge, \tool is the first to employ a generate-verify-refine paradigm for LLM-based ranking function synthesis.
\tool formulates ranking function synthesis as a constrained code-completion task, inserting placeholders into the input program that the LLM must fill in four predefined slots per loop. The resulting program is checked by a symbolic verifier. Successful verification yields a valid ranking function, while verification failures produce diagnostics that are translated into natural-language feedback to guide iterative refinement.
We evaluate \tool on two public benchmark suites comprising 358 terminating C programs. Using DeepSeek-V4.1-Flash, \tool synthesizes verified ranking functions for 333 programs (93.02\%), outperforming all evaluated baselines. Notably, it solves 22 more programs than {\sf PROTON}~2.1, the winner of the SV-COMP 2026 Termination Category. Combining \tool with {\sf PROTON}~2.1 further increases the success rate to 96.65\% (346/358 programs). Ablation studies demonstrate the contributions of placeholder contracts, few-shot examples, symbolic verification, and diagnostic feedback, while experiments across different LLMs establish the broader applicability of the framework.
\end{abstract}


\section{Introduction}

\IEEEPARstart{P}{rogram} termination is a fundamental liveness property that is essential for software reliability and total correctness~\cite{Floyd1967}, and plays an important role in software verification~\cite{DSilvaKW08,ShiXLZCL22,AlonD22}. 
While termination is undecidable in general~\cite{turing1936computable}, numerous automated techniques have been developed to establish termination for various classes of programs. A standard approach is to synthesize a ranking function that maps program states to a well-founded set and strictly decreases with each loop iteration. 
The existence of such a function guarantees that the loop terminates, as an infinite sequence of strictly decreasing values in a well-founded set is impossible.

Classical symbolic techniques for ranking function synthesis typically reduce the problem to constraint solving over predefined parameterized templates, such as linear expressions, lexicographic orders, or multiphase combinations~\cite{BenAmramG17,ben2013linear,bradley2005termination,larraz2013proving,LeikeH15,podelski2004complete,PodelskiR05}. 
A satisfying assignment to the template parameters yields a valid ranking function. While these approaches have achieved considerable success, their effectiveness depends heavily on the choice of templates. 
Designing suitable templates for diverse loop programs remains challenging and is still an active research area (cf., e.g.~\cite{LeikeH15}). 
Consequently, classical approaches often require substantial manual effort or computationally expensive template exploration, which limits their scalability and degree of automation.

To address these limitations, recent work has explored leveraging Large Language Models (LLMs) to generate ranking functions, motivated by the remarkable capabilities in software engineering~\cite{HouZLYWLLLGW24}.
{\sf LOOPY}~\cite{kamath2024leveraging} prompts LLMs to generate loop invariants and candidate ranking functions directly from natural-language definitions, and then identifies valid candidates using Frama-C~\cite{BaudinBBCKKMPPS21}. 
In contrast, {\sf PROTON} 2.1~\cite{mukhopadhyay2025proton} fine-tunes the Llama-3.2-1B-Instruct model to generate candidate ranking functions.
Despite their different approaches, both tools fall substantially short of established symbolic techniques in the number of programs for which they successfully synthesize valid ranking functions. This performance gap suggests that direct LLM-based generation, even with expert knowledge, has not realized its full potential for ranking function synthesis.

There are two major challenges in effectively leveraging LLM for ranking function generation.
\begin{itemize}[left=5pt] 
    \item
{\bf Challenge 1: limited 
    LLM capability.} 
    LLMs generate outputs based on statistical patterns learned from training data.
    As training data consists of large-scale 
    natural language text and code corpora, LLMs typically 
    perform well on tasks such as natural language processing and code generation. In contrast, ranking functions are relatively underexplored and is not supported by large or representative training datasets. Consequently, LLMs may have limited capability in reasoning about ranking functions and often struggle with complicated or subtle definitions expressed in natural language. Indeed, out-of-the-box Llama-1B failed to synthesize valid ranking functions even for trivial programs~\cite{mukhopadhyay2025proton}.

 \item  
 {\bf Challenge 2: identifying and resolving incorrect outputs.} Despite the promising performance of LLMs for code generation, 
they frequently generate incorrect code, often due to limited reasoning context. A particularly serious issue is hallucinations~\cite{ZhangWWSMZCMZ25} where LLMs generate outputs that appear plausible but fail to satisfy the intended specification~\cite{hagendorff2024deception,Caoetal25}. For instance, GPT-4 has been shown to generate incomplete assertions as 
post-conditions so that buggy programs do not violate them~\cite{EndresFCL24}. We indeed observe similar behavior in our experiments, highlighting the development of techniques that can effectively identify and correct such subtle errors with minimal interaction overhead. To date, no generate–verify–refine paradigm has been studied for LLM-based ranking function synthesis.
\end{itemize}


\smallskip
\noindent{\bf Our work.}
We propose a neuro-symbolic framework, \tool (\underline{RA}nk\underline{IN}g \underline{F}unction gener\underline{A}tion with \underline{LL}m), that, to our best knowledge, is
the first to adopt a generate-verify-refine paradigm for LLM-based ranking functions synthesis. \tool combines the generative capabilities of modern LLMs with the rigor of symbolic techniques. This integration enables \tool\ to discover richer termination arguments and prove termination for a broader class of loop programs. 
Unlike {\sf PROTON} 2.1, \tool requires no task-specific LLM fine-tuning,
avoiding its computational overhead and dependence on
high-quality training data. Moreover, fine-tuning on limited datasets may lead to overfitting and poor generalization to unseen programs.




To address the first challenge, 
we first transform the synthesis of ranking functions into a code-assertion generation task, namely, placeholder contracts, 
motivated by the observation that LLMs are highly effective at generating code and specifications.
This transformation 
improves the quality of the generated candidate ranking functions. 
Second, we embed expert knowledge about termination arguments (e.g., lexicographic and multi-phase ranking functions)  into structured prompts with few-shot examples, which can enhance the reasoning ability and in-context learning of LLMs. 


To address the second challenge, 
we present a novel symbolic verification pipeline that checks
whether the LLM outputs 
meet the required correctness criteria, thereby guaranteeing soundness. 
The pipeline comprises three checks: syntactic program-equivalence checking, syntactic code-assertion completeness checking, and semantic assertion checking.
The first one prevents potential cheating where the LLM modifies the original program behavior and generates assertions that hold only in the modified version.
The second one ensures that LLM-inserted code-assertion statements are complete (i.e., strict decrease, non-negativity and termination conditions),
The final one verifies whether the program satisfies the assertions via safety checking.
To repair incorrect outputs with minimal interaction, we re-prompt the LLM by incorporating 
fine-grained diagnostics (e.g., the detected program modification, missing ranking function conditions,
or counterexample produced by the safety verifier). 
A valid ranking functions can then be extracted 
and used as witness to program termination.

We evaluate \tool on two public benchmark suites comprising 358 terminating C programs. 
Using DeepSeek-V4.1-Flash~\cite{DeepSeek}, \tool generates verifiable ranking functions for 333 programs (93.02\%), 
outperforming seven state-of-the-art tools such as 
{\sf AProVE}~\cite{LommenG25} (78.77\%), {\sf UAutomizer}~\cite{HeizmannCDGHLNM18} (79.05\%),  {\sf LOOPY} (87.71\%) and  {\sf PROTON} 2.1 (86.87\%).
Ablation studies confirm that placeholder contracts, together with few-shot examples and diagnostic feedback is critical to this performance. We further show that \tool remains effective across LLMs (e.g., proved more programs in less time using GPT V5.5~\cite{GPT}) and complements other tools (e.g., a 96.65\% success rate, solving 346 tasks when combined with the symbolic counterpart of {\sf PROTON} 2.1).

\smallskip
\noindent{\bf Summary.} This paper makes the following contributions:
\begin{itemize}[left=5pt,topsep=0pt]
    \item A new perspective that formulates ranking-function synthesis as a code-assertion generation task, enabling a seamless integration of LLMs with sound symbolic techniques.
    \item A neuro-symbolic framework within the generate-verify-refine paradigm, which combines expert knowledge, few-shot examples, symbolic checking, and diagnostic feedback to guide generation 
    and ensure soundness.
    \item An extensive evaluation showing that \tool outperforms seven state-of-the-art tools including LLM-based 
    and top-3 tools of SV-COMP 2026 termination category. 
\end{itemize}


\smallskip
\noindent{\bf Structure.}  Section~\ref{sec:preliminaries} gives the background on program termination and common rank functions.
Section~\ref{sec:method} presents the \tool approach. Section~\ref{sec:evaluation} reports the experimental results 
and discusses potential threats to the validity of the work.  
We  discusses related work in Section~\ref{sec:related}
and conclude the paper in Section~\ref{sec:conc}.


\section{Background}
\label{sec:preliminaries}

\subsection{Program Termination}

A binary relation $R \subseteq W \times W$ over a set $W$ is \emph{well-founded} if every non-empty subset $W' \subseteq W$ has an 
$R$-minimal element, i.e., there exists $x' \in W'$ such that $(x, x') \not\in R$ for any $x \in W'$.
Intuitively, this means that no infinite descending chain with respect to $R$ exists. The relation $<$ over natural numbers is well-founded.
More generally, $<$ is well-founded over any finite set
of numerical values. 
In contrast, $<$ is not well-founded over the rational
or real numbers.
In this work, we denote by $\mathbb{K}$ a numerical
domain such that the relation $<$ is well-founded
over its nonnegative subset $\mathbb{K}_{\geq 0}$.


In this paper, we focus on programs with a single while-loop. Programs with multiple or nested loops can be handled by
reducing to a set of single loop programs~\cite{HeizmannHP14,BrockschmidtCIK16}, and for-loops can be easily transformed into while-loops,
thus are left as interesting future work.
The executions of such programs can be modeled as binary relations over program states.
Let $\vec{x}$ denote the vector of all program variables $(x_1, \ldots, x_n)^T$, representing the current program state, where $n \geq 1$.
(Note that the program variables may have different types.)
As a convention, we write $\vec{x}'=(x'_1, \ldots, x'_n)^T$ for the variables in the next program state.


\begin{definition}
    A program is modeled as a \emph{transition system} $\mathcal{P} := (S, I, R)$,
    where $S$ denotes the countable
(possibly infinite) set of reachable program states, $I \subseteq S$ represents the set of initial states, and $R\subseteq S\times S$ defines the transition relation over program states. 
\end{definition}

For program verification, the sets $S$, $I$, and $R$ are
typically represented as predicates over the program variables $\vec{x}$, yielding $I(\vec{x}), S(\vec{x})$ and $R(\vec{x}, \vec{x}')$, respectively. (We also identify these predicates with the corresponding sets of program states when the context is clear.)  
Intuitively, the initial states represent the reachable program states before entering the loop, and each pair $(s,s')\in R$ indicates that executing one (loop) iteration from the program state $s$ results in the program state $s'$. For the sake of presentation,
intermediate program states in each iteration are not modeled explicitly.

A \emph{computation} of the program $\mathcal{P}$ is either a finite or an infinite sequence $\pi = s_0, s_1, s_2, \cdots$ of program states
such that $s_0 \in I$ and $(s_i, s_{i+1}) \in R$ for all $i \geq 0$. A program $\mathcal{P}$ is \emph{terminating} if all its computations are finite.

A standard approach for proving termination of $\mathcal{P}$ is to synthesize a ranking function that maps program states to elements of a well-founded set such that its value strictly decreases along every transition. Since well-founded sets admit no infinite descending chains, the existence of such a ranking function guarantees that every computation of $\mathcal{P}$ is finite when the lower bound
entails the termination of the loop (i.e., negation of the loop guard).
Moreover, every (deterministic) terminating program has a
ranking function~\cite{Floyd1967}.

\subsection{Common Classes of Ranking Functions}
Practical methods for proving termination typically restrict the search to a specific class of ranking functions, otherwise the problem
is as hard as termination itself. We recall three common classes of ranking functions~\cite{LeikeH15}.
Note that the lower bound of the ranking functions considered below can be an arbitrary constant $c$ rather than $0$,
that is,  $\mathbb{K}_{\geq 0}$ can be generalized to $\mathbb{K}_{\geq c}$.
Prior work, e.g.,~\cite{LeikeH15}, typically fixes the lower bound to $0$; we adopt the term ``non-negativity'' to refer to the case where the lower bound is $0$.

\begin{definition}\label{def:srnkfunc}
A \emph{single ranking function} for a program
$\mathcal{P}=(S,I,R)$ is a function
$f:S\to\mathbb{K}$ such that for every
$(s,s')\in R$, both $f(s)\geq 0$ and 
$f(s)>f(s')$ hold.
\end{definition}

Intuitively, $f$ maps each program state $\vec{x}$ to a value $f(\vec{x})$ 
which strictly decreases after each iteration and is nonnegative whenever an iteration starts.
The ranking value may become negative after the final iteration.
Since the strict ordering over $\mathbb{K}_{\geq 0}$ is well-founded, the loop with guard $\phi$ can execute only finitely many iterations when $f(s)<0$ entails $\neg \phi$. Therefore, $f$ serves as a witness to program termination.

However, single ranking functions are not always sufficient, or easy to construct. 
For example, consider the program $\mathcal{P}$ in Figure~\ref{fig:example-program}.
For brevity, we omit (1) the declaration {\tt extern int \_\_VERIFIER\_nondet\_int(void)} which returns a nondeterministic integer, and (2) the effects of integer overflow and underflow when reasoning about the termination of $\mathcal{P}$. This follows from prior work~\cite{podelski2004complete,BradleyMS05,larraz2013proving,LeikeH15,BenAmramG17}, since termination is otherwise undefined. Such issues are typically addressed separately by safety verification.

We claim that $\mathcal{P}$ always terminates (assuming no integer overflow occurs). Intuitively, once $y$ becomes positive, $x$ strictly decreases in every subsequent iteration. Since $x$ cannot decrease indefinitely while satisfying the loop guard $x > 0$, the loop must eventually terminate. 
Nevertheless, there exists no single \emph{linear} ranking function of the form $f(x, y) =ax + by + c$,  where $a,b$ and $c$ are constants, that proves the termination of $\mathcal{P}$, since such a function may fail to strictly decrease at every iteration.

To see this, let the current program state be $(x, y)$.  After one iteration,
the program state is updated to $(x-y, y+1)$.
We have that $f(x, y) - f(x-y, y+1)= (ax + by + c) - (a(x-y) + b(y+1) + c)  = ay -b $.
\begin{itemize}[left=5pt, topsep=2pt]
    \item When $a \neq 0$, for $f$ to be a valid ranking function, it holds that $ay -b > 0$ for all program states satisfying the loop guard, including $y = \frac{b}{a}$ and $x = 2$. But then we have $ay-b = 0$, contradicting Definition~\ref{def:srnkfunc} since $f$ does not strictly decrease at every iteration.
    \item When $a = 0$, we require that 
    $-b \geq 0$ according to the strict decrease requirement Definition~\ref{def:srnkfunc}.
    It means that $b < 0$. However, for a fixed value $c$, we have $f(x, y) = by+c < 0$ when $y > -\frac{c}{b}$, which contradicts the requirement $f(x, y) \geq 0$ in Definition~\ref{def:srnkfunc}.
\end{itemize}
Hence, the termination of $\mathcal{P}$ cannot be proved by any single linear ranking functions.
In contrast, this program admits a multiphase ranking function. 


\begin{figure}[!t]
    \centering
    \begin{lstlisting}
int main() {
    int x = __VERIFIER_nondet_int();
    int y = __VERIFIER_nondet_int();
    // >>> Infill Define <<<
    while ( x > 0 ) {
        // >>> Infill Update Begin <<<
        x = x - y;
        y = y + 1;
        // >>> Infill Update End <<<
        // >>> Infill Assert <<<
    }
    return 0; }
\end{lstlisting}
\vspace{-2mm}
    \caption{An example (terminating) loop program $\mathcal{P}$.}
    \label{fig:example-program}\vspace{-4mm}
\end{figure}

\begin{definition} 
A \emph{multiphase ranking function} for a program
$\mathcal{P}=(S,I,R)$ is a tuple
$\vec{f}=(f_1,\ldots,f_k)$ of component functions
$f_i:S\to\mathbb{K}$ such that, for every
$(s,s')\in R$, there exists $i\in\{1,\ldots,k\}$
satisfying (1) $f_i(s)\geq 0$ and $f_i(s)>f_i(s')$, and (2)
$f_j(s)<0$ and $f_j(s')<0$ hold for all $j<i$.
The component $f_i$ is said to be active in this transition and function $\vec{f}$ is referred to as a
\emph{$k$-phase ranking function}.
\end{definition}

As the active component strictly decreases while nonnegative,
completed components remain negative, and the number of phases
is finite, multiphase ranking functions can witness program
termination with the loop guard $\phi$ when
$f_k(\vec{x}) < 0$ entails $\neg \phi$.
For instance, consider the 2-phase ranking function
$\vec{f}(x,y) = (-y, x)$ for the program $\mathcal{P}$
in Figure~\ref{fig:example-program}.
The first component $-y$ strictly decreases after every
iteration because $y$ is incremented by one,
while the second component $x$ may increase during this phase.
Once $y > 0$ (i.e., $-y < 0$), the first component
remains negative, while the second component $x$
strictly decreases until the loop eventually terminates.

We now introduce another common class of ranking functions, i.e., lexicographic ranking functions.

\begin{definition} 
A \emph{lexicographic ranking function} for a program
$\mathcal{P}=(S,I,R)$ is a tuple
$\vec{f}=(f_1,\ldots,f_k)$ of component functions
$f_i:S\to\mathbb{K}$ such that, for every
$(s,s')\in R$, the following conditions hold:
(1) $f_i(s)\geq 0 $ for all $1\leq i\leq k$,
and (2) there exists $i\in\{1,\ldots,k\}$ satisfying
$f_i(s)>f_i(s')$, and $f_j(s)=f_j(s')$ for all $j<i$.
Such a function $\vec{f}$ is referred to as a
\emph{$k$-lexicographic ranking function}.
\end{definition}


Since the lexicographic order over $\mathbb{K}_{\geq 0}^k$
is well-founded, a strictly decreasing lexicographic ranking
function witnesses program termination with the loop guard
$\phi$, provided that
$\bigvee_{i=1}^k (f_i(\vec{x})<0)$ entails $\neg\phi$.
For instance, consider the 2-lexicographic ranking function
$\vec{f}(x,y) = (y \leq 0\ ? 1-y : 0, x)$
for the program $\mathcal{P}$ in
Figure~\ref{fig:example-program}.
When $y\leq 0$, the first component is $1-y$ and
strictly decreases after every iteration, because $y$
is incremented by one.
Once $y>0$, the first component remains $0$,
and the second component $x$ strictly decreases
until the loop eventually terminates.

\begin{figure*}[!t]
  \centering
  \includegraphics[width=.99\textwidth]{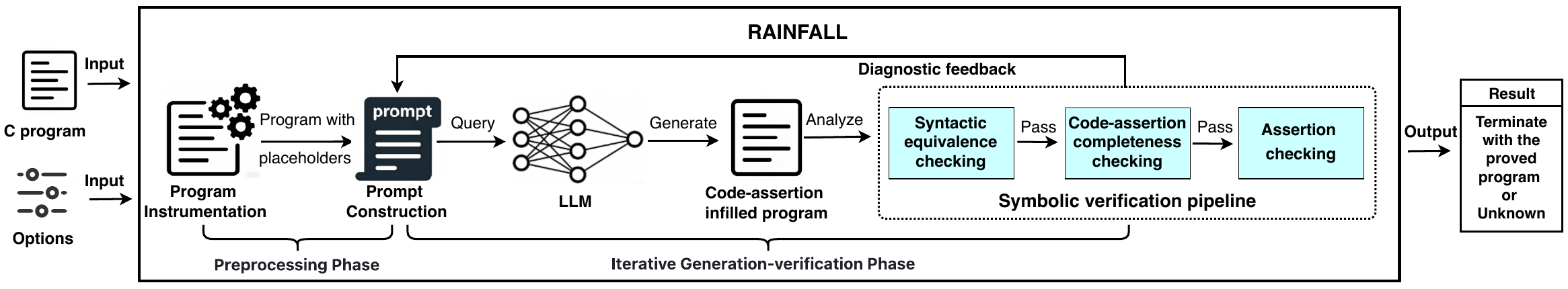}\vspace{-2mm}
  \caption{An overview of \tool.}\vspace{-4mm}
  \label{fig:overview}
  \end{figure*}

\section{Our Approach}
\label{sec:method}

Figure~\ref{fig:overview} illustrates the overall workflow of \tool.
Given a C program $\mathcal{P}$ and resource constraints (e.g., budgets on LLM queries and execution time), \tool aims to synthesize a ranking function $f$ within these limits to prove
the termination by generating a verified program $\mathcal{P}_{\tt proved}$. If successful, \tool outputs {\tt Terminate} along with the LLM-generated and symbolically verified program $\mathcal{P}_{\tt proved}$  as a certificate of termination; otherwise it returns {\tt Unknown}.

\tool consists of two main phases: \emph{preprocessing} and \emph{iterative generation-verification}. In the preprocessing phase (cf.~Section~\ref{sect:pre}), \tool performs \emph{program instrumentation} and 
\emph{prompt construction}. It first automatically instruments $\mathcal{P}$ by inserting four predefined code and assertion placeholders at selected locations for the loop, resulting in an instrumented program $\mathcal{P}_{\tt instr}$.
This reduces ranking function synthesis to a code-assertion generation problem
on $\mathcal{P}_{\tt instr}$. Next, 
it constructs a structured prompt $\mathcal{M}_{\tt init}$ based on $\mathcal{P}_{\tt instr}$,  natural language descriptions and few-shot examples of ranking functions, and queries the LLM accordingly. 

In the iterative generation-verification phase (cf.~Section~\ref{sect:ite}), \tool generally alternates between
\emph{symbolic verification} and \emph{prompt construction} with diagnostic feedback.
It first queries the LLM with the initial prompt $\mathcal{M}_{\tt init}$ (from the preprocessing phase), to produce a candidate program $\mathcal{P}_{\tt cand}$ 
by infilling the code/assertion placeholders. 
Despite the structured prompt, 
$\mathcal{P}_{\tt cand}$ may still be incorrect or incomplete, and thus is 
analyzed by a symbolic verification pipeline. 

The symbolic verification pipeline comprises three components: 
\emph{syntactic equivalence checking}, which ensures that the original program behavior in $\mathcal{P}$ is not modified by the LLM; \emph{code-assertion completeness checking}, which verifies that the generated assertions cover strict decrease, non-negativity, and termination conditions; 
and \emph{assertion checking}, which validates the correctness of these assertions. 
If $\mathcal{P}_{\tt cand}$ passes all three checks, the LLM-generated code-assertions induce a valid ranking function $f$, 
and \tool returns $\mathcal{P}_{\tt cand}$ as $\mathcal{P}_{\tt proved}$. Otherwise, diagnostics are incorporated into the
$\mathcal{M}_{\tt init}$, leading to the prompt $\mathcal{M}_{\tt fb}$ for the subsequent iteration (a new interaction with the LLM using $\mathcal{M}_{\tt fb}$), to refine the LLM output. 
This process repeats until a valid ranking function is found, or the resource budget is exhausted.  



\begin{figure*}[!t]
  \centering
  \includegraphics[width=1\textwidth]{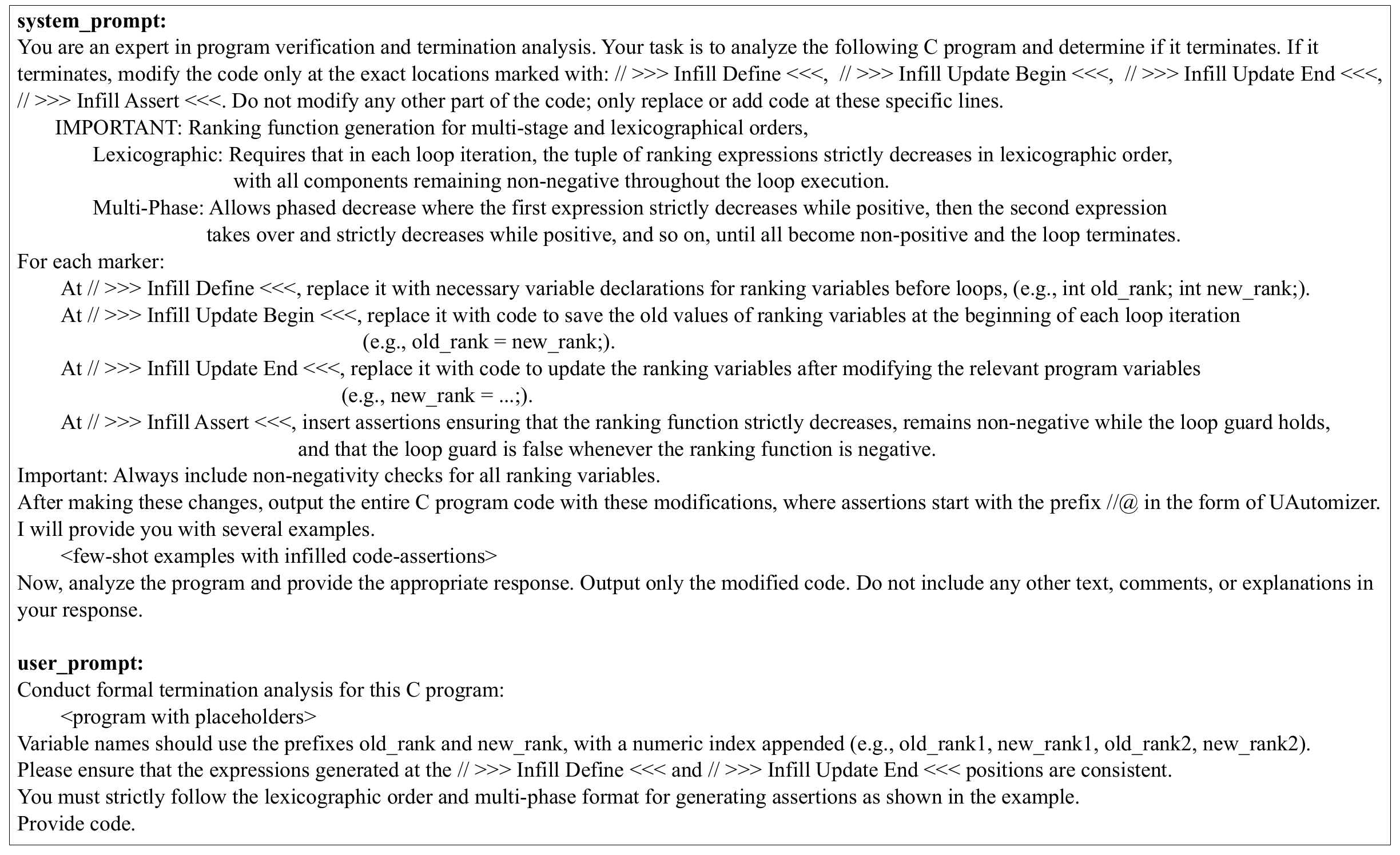}  \vspace{-5mm}
  \caption{Initial prompt template.}
  \label{fig:prompt-demo}\vspace{-4mm}
  \end{figure*}


\subsection{Preprocessing} \label{sect:pre}
The goal of the preprocessing phase is to instrument the input program $\mathcal{P}$ with placeholders to be infilled by the LLM, and to build an initial prompt that guides the LLM in filling the placeholders so as to obtain a valid ranking function $f$.

\begin{figure*}[!t]
  \centering
  \includegraphics[width=1\textwidth]{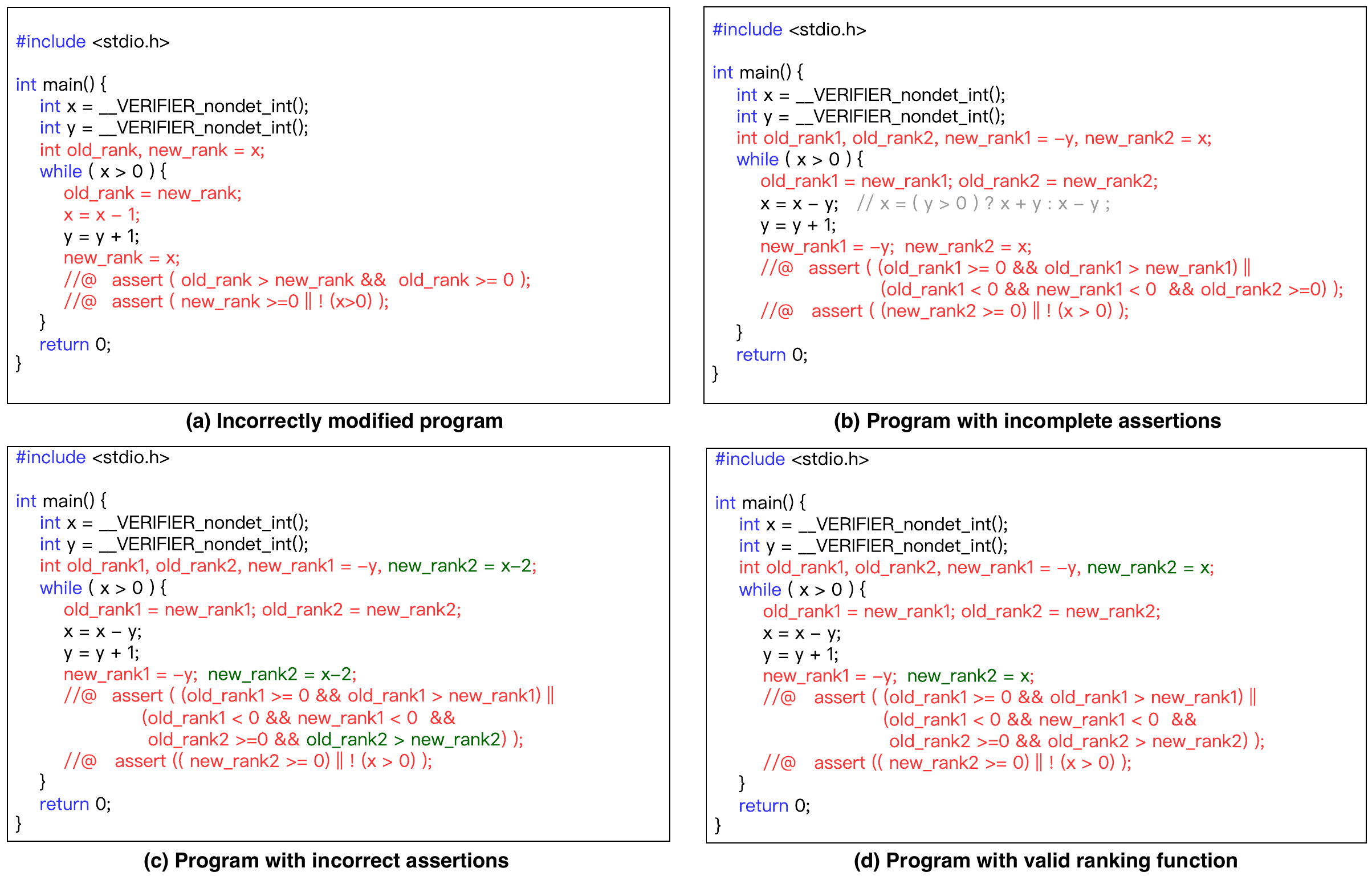}  \vspace{-5mm}
  \caption{Demonstration of \tool on the program in Figure~\ref{fig:example-program}.}
  \label{fig:demo}\vspace{-2mm}
  \end{figure*}

\smallskip
\noindent\textbf{Program Instrumentation.}
We first construct the abstract syntax tree (AST) of the program $\mathcal{P}$
and identify its while-loop 
by traversing AST. Recall that in this work, we focus on programs with a single while-loop, cf. Section~\ref{sec:preliminaries}). 
Then, all the {\tt continue} statements are replaced by the statement {\tt goto loop\_exit}
and the fresh label {\tt loop\_exit} is inserted at the end of the loop if it does not exist. Note that {\tt loop\_exit} can be any other label if 
it is used elsewhere. Let $\mathcal{T}$ be the resulting AST and regard it as the original program.
Next, we insert four predefined placeholders at the designated locations, which are to be infilled by the LLM with code and assertions.
More precisely, as illustrated in Figure~\ref{fig:example-program} and Figure~\ref{fig:demo}(a), we introduce the following four placeholders for the loop:

\begin{itemize}[left=5pt, topsep=2pt]
    \item {\tt // $>>>$ Infill Define $<<<$}: Placed before the loop to declare auxiliary variables required by the ranking function (e.g.,  variables \verb|old_rank| and  \verb|new_rank| are declared to record the ranking value from the previous iteration and the current ranking value, i.e., before and after each loop iteration). Such auxiliary variables are referred to as \emph{ghost variables}.
    \item {\tt // $>>>$ Infill Update Begin $<<<$}: Placed at the beginning of the loop body to insert statements that store the current ranking value before the start of the current iteration (e.g., 
    {\tt old\_rank=new\_rank} assigning the current value to the variable \verb|old_rank| representing the previous value).
    \item {\tt // $>>>$ Infill Update End $<<<$}: Placed at the end of the loop body to insert statements that update the current ranking value after the current loop iteration 
    (e.g.,  \verb|new_rank=x| updates \verb|new_rank| according to the value of the variable \texttt{x}).
    \item {\tt // $>>>$ Infill Assert $<<<$}: Placed at the end of the loop body (before the label {\tt loop\_exit} if it exists) to insert assertions that enforce the strict decrease and non-negativity of the ranking value, as well as termination of the loop (e.g., \verb|assert(old_rank>new_rank)| enforces the strict decrease).
\end{itemize}

By these placeholders, we restrict the LLM to act as a constrained slot filler rather than a free-form code generator, aiming to prevent unintended modifications to other parts of the program logic and enhance the predictability and reliability of the generated results. The resulting instrumented program is denoted as $\mathcal{P}_{\tt instr}$.

\smallskip
\noindent\textbf{Initial Prompt Construction.}
To guide the LLM in correctly infilling the placeholders,
as illustrated in Figure~\ref{fig:prompt-demo},
we construct a structured prompt $\mathcal{M}_{\tt init}$, consisting of the following four key parts:

\begin{itemize}[left=5pt, topsep=2pt]
    \item \emph{System Role}: Specifies that the current task is to synthesize a ranking function for termination analysis of C programs by filling the four predefined placeholders for the loop. It furthermore emphasizes strict  
    placeholder replacement rules: only the predefined placeholders may be replaced, and all other parts of the program must remain unchanged.

    \item \emph{Semantic Mode Definition}: Encodes domain knowledge (e.g., decreasing conditions for lexicographic and multiphase ranking functions) in natural language, guiding the LLM to correctly capture the semantics of complex termination arguments.

    \item \emph{Few-shot Examples}: Provides fully instantiated programs, 
    including variable declarations, update statements and assertions, to illustrate the expected output format and implicitly convey common patterns of ranking function synthesis via  code-assertion filling, thereby improving generation accuracy.

   \item \emph{Instrumented Program}: The placeholder-instrumented program $\mathcal{P}_{\text{instr}}$ produced in the preprocessing phase, serving as the concrete context for generation.
\end{itemize}

The resulting prompt $\mathcal{M}_{\tt init}$ is used to query the LLM, which often
generates a valid ranking function by filling the placeholders. However, incorrect or incomplete generations may still arise; such cases will be handled by the subsequent iterative generation-verification
phase.

\subsection{Iterative Generation-Verification} \label{sect:ite}

This phase repeatedly queries the LLM and verifies
the LLM-generated candidate program $\mathcal{P}_{\tt cand}$ via a designated symbolic verification pipeline,
until $\mathcal{P}_{\tt cand}$ satisfies the required correctness criteria
or the resource budget is exhausted.
This process begins by querying the LLM using $\mathcal{M}_{\tt init}$ and iteratively 
refines the LLM output by incorporating diagnostic feedback extracted from the symbolic verification pipeline into the prompt for the subsequent iteration.

We first describe the three components of the symbolic verification pipeline, 
namely, \textbf{syntactic equivalence checking}, \textbf{code-assertion completeness checking} and \textbf{assertion checking}. We further explain how their diagnostic feedback is incorporated into the prompt to refine the LLM output.

\subsubsection{Syntactic Equivalence Checking}
While the prompt $\mathcal{M}_{\tt init}$ explicitly restricts the LLM to be a slot filler and disallows modifications to the program logic, our experiments show that the LLM may still modify the code in $\mathcal{P}_{\tt instr}$. 

For instance, when querying the LLM 
on the instrumented program
in Figure~\ref{fig:example-program}, the model may generate a candidate ranking function by filling the four placeholders, as shown in Figure~\ref{fig:demo}(a).
We highlight the instantiated placeholders and the modified statement in \textcolor{red}{red}.
In the first slot, two variables \verb|old_rank| and \verb|new_rank| are declared to record the previous and current ranking values, respectively, initializing 
\verb|new_rank| with {\tt x}. In the second slot, the assignment \verb|old_rank=new_rank| is inserted to store 
the ranking value before the iteration. 
In the third slot, the ranking value is updated after the loop body via \verb|new_rank=x|.  
In the fourth slot, two assertions are inserted: the first enforces  
strict decrease and non-negativity (i.e., \verb|old_rank>new_rank| and \verb|old_rank>=0|), while
the second encodes termination by asserting that if the ranking value becomes negative, the loop guard must be false (i.e., \verb+(new_rank>=0)||!(x>0)+).  

These code-assertions indeed induce a single ranking function  $f(x,y)=x$, which is valid
for the program in Figure~\ref{fig:demo}(a), and would therefore pass the subsequent 
completeness and assertion checks.
However, the statement \verb|x=x-y| in the original program code 
is incorrectly replaced by \verb|x=x-1|, meaning that $f$ is \emph{not} a valid
ranking function for the original program.

To detect such incorrect outputs, we adopt a lightweight \emph{syntactic} equivalence checking to ensure that the original code in $\mathcal{P}$ is not modified by the LLM. 
We build ASTs for both the original program $\mathcal{P}$ and the LLM-generated program $\mathcal{P}_{\tt cand}$ (after removing the code-assertions inserted by the LLM), and compare them for equivalence. $\mathcal{P}_{\tt cand}$ proceeds to the next checking if it passes this equivalence check; otherwise, 
diagnostics are fed back to refine
the LLM output.


We do not consider \emph{semantic} equivalence checking as it is computationally expensive and, in our setting, LLMs rarely generate semantically equivalent programs when the program code is modified. Moreover, we do not embed LLM-generated code-assertions into the original program $\mathcal{P}$, since such code-assertions typically cannot 
induce a valid ranking function for $\mathcal{P}$ and subsequent assertion checking would incur substantial computational overhead. 

Finally, we note that $\mathcal{P}_{\tt instr}$ may fail to preserve
the original program behavior even when it passes our syntactic equivalence check. For example, updating a ghost variable may change program behavior through side effects. Although this never occurred in our experiments, such cases should still be checked to ensure soundness.
To ensure that LLM-inserted assignments do not alter the original program behavior, 
we parse all the expressions assigned to ghost variables in the first and third slots and check that they are pure expressions built from 
program variables and constants, containing no function calls, no assignments to program variables, and no increment/decrement expressions on program variables.
If an expression fails this check, 
a grammar-violation diagnostic is returned to the LLM and the candidate program is not passed to subsequent checks.
In addition, our assertion completeness checking implicitly ensures that LLM-inserted assertions do not alter the original program behavior.

\subsubsection{Code-Assertion Completeness Checking.}
%
As observed in our experiments, the LLM may generate incomplete code-assertions. 
For instance, when querying the LLM 
on the instrumented program in Figure~\ref{fig:example-program}, the model may generate the candidate ranking function shown in Figure~\ref{fig:demo}(b). 
In the first slot, four variables \verb|old_rank1|, \verb|old_rank2|, \verb|new_rank1|
and \verb|new_rank2| are declared, with \verb|new_rank1| and \verb|new_rank2| initialized to $-y$ and $x$, respectively. 
The second and third slots update these variables.
In the fourth slot, two assertions are inserted to encode a 2-phase ranking function $(f_1,f_2)$, where the first enforces the strict decrease and non-negativity 
of $f_1$ (with previous and current values stored in \verb|old_rank1| and \verb|new_rank1|),
and, once $f_1$ becomes negative, enforces 
non-negativity of $f_2$ (with 
values stored in \verb|old_rank2| and \verb|new_rank2|); the second assertion encodes termination when $f_2$ becomes negative.

At first glance, these assertions appear to induce the following 2-phase ranking function 
\[
f(x,y)=(f_1(x,y),f_2(x,y))=(-y,x)
\]
and would pass the assertion check.
However, the first assertion is incomplete: it fails to enforce strict decrease of $f_2$ after $f_1$ becomes negative. As a result, the non-terminating program obtained by replacing \verb|x=x-y| with \verb|x=(y>0)?x+y:x-y|
still satisfies both assertions, as $-y$ strictly decreases 
while $x$ strictly increases 
when $y\neq 0$.

To detect such incomplete outputs, we perform code-assertion completeness checking as follows:

%

\smallskip
\noindent
{\bf Checking the first slot.}
We check the completeness of the \emph{ghost variable} declarations in the first slot. We identify all declared ghost variables by parsing the declarations and partition them into two sets $G=\{g_1,\cdots, g_n\}$ and $G'=\{g_1',\cdots, g'_m\}$,  
depending on whether they are initialized or not. Here, $G$ represents the components of the previous ranking value (i.e., before the start of the current iteration), while $G'$ represents the components of the current ranking value ($g_i'$ is initialized with $e_i$ for $1\leq i\leq m$). We require that there is a one-to-one correspondence between $G$ and $G'$, so we check whether they have the same cardinality,
i.e., $m=n$; otherwise, the declarations are incomplete. W.l.o.g., we assume that
for each $1\leq i\leq m$, $g_i$ and $g_i'$ record the $i$-th component of the ranking value, established by
the one-to-one correspondence.

\smallskip
\noindent
{\bf Checking the second slot.}
We then check the completeness of the statements in the second slot by parsing the filled statements. These statements should assign the previous values from the current ones before the current loop iteration, i.e., $g_i = g_i'$ for all $1 \leq i \leq n$; otherwise, the updates at the beginning of the loop body are incomplete.
 
\smallskip
\noindent
{\bf Checking the third slot.}
Next, we check the completeness of the statements in the third slot by parsing the filled statements. These statements should update the ranking values after executing the current loop iteration, i.e., $g_i' = e_i$ for all $ 1\leq i \leq k$,
where $e_i$'s are the same expressions as in the first slot. Otherwise, the updates at the end of the loop body are incomplete.

\smallskip
\noindent
{\bf Checking the fourth slot.}
Finally, we check the completeness of the assertions, ensuring that they capture the strict decrease, non-negativity and termination conditions required by the target class of ranking functions. This checking is much involved and
depends on the classes of ranking functions.
Let \verb|bc| denote the loop guard. 
\begin{itemize}[left=5pt, topsep=2pt]
    \item \emph{Simple ranking function:} To induce a simple ranking function, there should be only two ghost variables $g_1$ and $g_1'$, namely, $n=m=1$. Therefore, we check if LLM-generated outputs assert the following three predicates: 

    \begin{center}
    \begin{tabular}{ll}
      Strict decrease condition: & $g_1 > g_1'$ \\
      Non-negativity condition: & $g_1 \geq 0$ \\
      Termination condition: & $g_1' \geq 0 \lor ! bc$
    \end{tabular}.
    \end{center}
    
For example, the assertions in Figure~\ref{fig:demo}(a)
with the loop guard $x>0$ conform to the these predicates.  
Thus,  it is complete for a single ranking function.
    \item \emph{Multiphase ranking function:} It can only induce a $k$-phase ranking function with $k=n>1$. 
    Thus, we check if LLM-generated outputs assert the following two predicates
    (strict decrease and non-negativity, and termination):
    
    \begin{tabular}{c}
         $\left\{\begin{array}{c}
       ( g_1\geq 0 ~\&\&~  g_1>g_1' ) ~|| \\
       ( g_1< 0  ~\&\&~ g_1'< 0~\&\&~  g_2\geq 0 ~\&\&~  g_2>g_2' ) ~|| \\
         \cdots \\
       \left(\begin{array}{c} 
        g_1< 0~\&\&~ g_1'< 0 ~\&\&~  \\
       \cdots \\
       g_{n-1}< 0~\&\&~g'_{n-1}< 0  ~\&\&~ \\
       g_{n}\geq 0 ~\&\&~  g_n>g_n' 
        \end{array}\right)
      \end{array}\right\}$  \\
      $g_n'\geq 0~ ||~ ! bc$
    \end{tabular}.

For example, the assertions in  Figures~\ref{fig:demo}(c) and~\ref{fig:demo}(d)
conform to the these predicates under the loop guard is $x>0$ and $n=m=2$, whereas those in Figure~\ref{fig:demo}(b) do not:
%
%
Thus, the programs in  Figure~\ref{fig:demo}(c) and Figure~\ref{fig:demo}(d) are complete for a $2$-phase ranking function
while the other one is not.


    \item \emph{Lexicographic ranking function:} It can only induce a $k$-lexicographic ranking function with $k=n>1$. 
    Therefore, we check if LLM-generated outputs assert the following three predicates (strict decrease,  non-negativity, and termination):

    \begin{tabular}{c}
       $\left\{\begin{array}{c}
       (g_1>g_1') ~||  \\ 
       (g_1==g_1' ~\&\&~ g_2>g_2') ~||  \\ 
         \cdots \\
        (g_1==g_1' \&\& \cdots \&\& g_{n-1}==g_{n-1}' \&\& g_n>g_n') 
      \end{array}\right\}$  \\
      $g_1>=0 ~\&\&~\cdots ~\&\&~ g_n>=0$ \\ 
      $(g_1'>=0 ~\&\&~\cdots ~\&\&~ g_n'>=0)~ ||~ ! bc$ 
    \end{tabular}
    
      We note that the strict decrease and termination conditions are sufficient 
      for lexicographic ranking functions to be complete, because if the loop does not terminate,
      then $bc$ always holds (i.e., $!bc$ is false), the condition $(g_1'>=0 ~\&\&~\cdots ~\&\&~ g_n'>=0)$ 
      will eventually be violated. We require LLMs to also generate the non-negativity condition
      so that our prompts are independent upon the classes of ranking functions.
\end{itemize}
  

If the LLM-generated assertions do not conform to any of these forms, they are deemed incomplete, and diagnostic feedback is generated and incorporated into the prompt. Otherwise, $\mathcal{P}_{\tt cand}$ proceeds to assertion checking.


\subsubsection{Assertion Checking.}
Both the equivalence checking and completeness checking are purely syntactic. 
To ensure soundness, a \emph{semantic} assertion checking is therefore required.

For instance, when querying the LLM 
on the instrumented program in Figure~\ref{fig:example-program}, the model may generate the candidate ranking function in Figure~\ref{fig:demo}(c). 
This program is similar to the one in Figure~\ref{fig:demo}(b), aiming to encode a 2-phase ranking function 
$f(x,y)=(-y,x-2)$ instead of $f(x,y)=(-y,x)$, where
(1) \verb|old_rank2>new_rank2| is inserted in the first assertion to enforce strict decrease of $f_2$ once $f_1$ becomes negative,
and (2) \verb|old_rank2 = x| is changed to \verb|old_rank2 = x-2|.
(These changes are highlighted in \textcolor{green}{green}.)
This program passes both the equivalence and code-assertion completeness checks. However,
although $f(x,y)=(-y,x-2)$ generally is a 2-phase ranking function, 
the components of the ranking value are not lower-bounded by $0$,
thus is not a valid ranking function for the original program in Figure~\ref{fig:example-program}.

Semantic assertion checking is thus necessary to verify the LLM-generated code-assertions as a valid ranking function. Specifically, the program $\mathcal{P}_{\tt cand}$ is verified using a safety verifier (e.g., {\sf UAutomizer}).
Intuitively, this step checks whether the original program $\mathcal{P}$ terminates under the candidate ranking function induced by the code-assertions in $\mathcal{P}_{\tt cand}$.
This procedure is sound because (1) the equivalence checking ensures that 
the original program behavior of $\mathcal{P}$ is not modified in $\mathcal{P}_{\tt cand}$, and (2) the completeness checking ensures that all the required code-assertions 
have been properly instantiated in $\mathcal{P}_{\tt cand}$. 
Consequently, if the assertions are valid in $\mathcal{P}_{\tt cand}$, they collectively induce a valid ranking function for
proving termination of $\mathcal{P}$. 

If $\mathcal{P}_{\tt cand}$ passes this check, \tool outputs {\tt Terminate} together with the program $\mathcal{P}_{\tt cand}$, which serves as a certificate of termination for
$\mathcal{P}$. Otherwise, a counterexample is typically reported by the safety verifier and 
fed back to improve the prompt.
For instance, {\sf UAutomizer} reported {\tt SAT} when verifying the program in Figure~\ref{fig:demo}(c), returning the violated
assertion together with a counterexample, i.e., a sequence of statements leading to the violated assertion, along with the corresponding program states. In this case, the input is $(x=1, y=1)$ and the program executes $$(x=1, y=1)\rightarrow (x=0, y=2),$$ 
with corresponding ranking values $$(-y=-1,x-2=-1),(-y=-2,x-2=-2).$$
We observe that although the second assertion is not violated, the first assertion is violated at the state $(x=0, y=2)$ with ranking value $(-y=-2,x-2=-2)$, since neither \verb|old_rank1 >=0| nor \verb|old_rank2 >=0| holds.

When querying the LLM with the counterexample augmented prompt $\mathcal{M}_{\tt fb}$, the model generates the program shown in Figure~\ref{fig:demo}(d). This program passes all three checks and induces
the valid 2-phase ranking function $f(x,y)=(-y,x)$, which is now sufficient to prove termination of $\mathcal{P}$. 
The key difference is that the second component in the ranking function is changed 
from $x-2$ to $x$ so that it is lower-bounded by $0$
and the termination assertion enforces that
\verb|new_rank2 < 0| entails \verb|!(x>0)|.
Consequently, \tool outputs {\tt Terminate} along with the verified program $\mathcal{P}_{\tt proved}$, thereby completing the ranking function generation and verification task.

\subsubsection{Prompt Augment with Diagnostic Feedback.}
To address the issues identified by the checks, we augment  
the initial prompt $\mathcal{M}_{\tt init}$
with the incorrectly generated program $\mathcal{P}_{\tt cand}$ and the corresponding diagnostics.
This combined text forms a complete context for the next new interaction with the LLM. 
The diagnostics are tailored to each type of check. For syntactic equivalence checking, it reports mismatched statements with their locations
or LLM-inserted expressions that change the original program behavior.
For code-assertion completeness checking, it highlights the incorrect/incomplete variable declarations and code within each placeholder slot.
For assertion checking, it provides 
the location of the violated assertion along with a counterexample, 
as reported by the underlying safety verifier.

For example,  equivalence checking for the program in Figure~\ref{fig:demo}(a) may produce a diagnostic such as ``the statement x=x-1 at location 10 in the generated program mismatches with the statement x=x-y at location 10 in the original program.''
The completeness checking for the program in Figure~\ref{fig:demo}(b) may report: 
``the first assertion misses the strict decreasing condition for the variables old\_rank2 and new\_rank2.''
For the program in Figure~\ref{fig:demo}(c), assertion checking may yield: ``We used UAutomizer, and it found a counterexample demonstrating that our program is incorrect. CounterExampleResult [Line: xx]: assertion can be violated.
We found a FailurePath: $\cdots$''.
If the safety verifier fails to prove or disprove the assertions (e.g., it reports {\tt Unknown} 
or runs out of execution time), the feedback is ``We used UAutomizer which reports Unknown/runs out of time.''

\section{Evaluation}
\label{sec:evaluation}

We implement \tool using Clang~\cite{clang} 
for AST construction and manipulation, and {\sf UAutomizer} as the backend safety verifier. \tool supports configurable settings, including the choice of LLM and the maximum number of generation-verification-refinement iterations. We systematically evaluate \tool to address the following research questions (\textbf{RQs}):

\begin{itemize}[left=5pt, topsep=2pt]
    \item \textbf{RQ1} (Effectiveness): How does \tool compare with state-of-the-art termination analysis tools? 
    \item \textbf{RQ2} (Ablation study): What are the individual contributions of placeholder contracts, few-shot examples, symbolic verification
    and diagnostic feedback?
    \item \textbf{RQ3} (Impact of LLMs): How robust is \tool across different LLMs?
\end{itemize}

\smallskip 
\noindent\textbf{Benchmark.}
We consider two public benchmark suites provided by {\sf LOOPY} and {\sf FreqTerm}~\cite{FedyukovichZG18}, denoted by BS-L and BS-F, respectively. 
BS-L consists of 188 terminating C programs after removing 93 duplicates with identical ASTs
but different line numbers. 
BS-F contains 170 terminating C programs after
removing one non-terminating program.
(Note that no deduplication is needed for BS-F, although
one program contains a for-loop that we transform
into an equivalent while-loop.)
%
Both suites were originally collected from the termination category of SV-COMP, covering typical loop patterns such as linear, non-linear and conditional branches. 
We retain overlapping programs across the two
suites to maintain compatibility with prior evaluations,
as some baselines, including \textsf{ddlterm}~\cite{xu2022data}
and {\sf LIFT}~\cite{gong2026llmguided},
support only BS-F.
The few-shot examples used in our experiments
do not overlap with the benchmark programs.


Due to the high token costs of LLM-based experiments,
we do not evaluate on the full SV-COMP Termination
benchmark set. Instead, we use these two established
suites, comprising 358 terminating C programs in total.
Their use is consistent with prior studies:
{\sf ddlterm} (ICSE 2022),
{\sf LIFT} (ICML 2026) and
{\sf CondTerm}~\cite{huang2026accurate}
(ICSE 2026) evaluate exclusively on BS-F
or its 170 terminating programs.
We use both BS-L and BS-F to answer
\textbf{RQ1}, while restricting 
\textbf{RQ2} and \textbf{RQ3} to BS-F to control the token costs of
additional experiments.

\smallskip 
\noindent\textbf{Baselines.}
We consider seven representative baselines:
{\sf UAutomizer}, {\sf ddlTerm}, {\sf FreqTerm},
{\sf AProVE}, 
{\sf PROTON}, 
{\sf LOOPY}
and {\sf LIFT}. 
These tools represent a diverse range of approaches for termination proving (cf. Section~\ref{sec:related}).
Notably, {\sf PROTON}, {\sf UAutomizer} and {\sf AProVE} are 
the top three performers in the Termination Category of SV-COMP 2026~\cite{BeyerS26}.
We additionally consider the fine-tuned LLM of {\sf PROTON} (denoted as {\sf PRO-LLM})
and \toolv, a variant of \tool, which does not use code-assertion contract. 
\toolv prompts the LLM with expert knowledge and few-shot examples to directly generate 
a ranking function class and its corresponding expression(s). 
It then encodes the generated ranking function as code and assertions according to one of the three common classes of ranking functions, and uses the safety verifier {\sf UAutomizer} to validate. If verification fails, diagnostic feedback guides iterative refinement of the LLM's output. 

\smallskip 
\noindent\textbf{LLMs.}
For \textbf{RQ1} and \textbf{RQ2}, we use DeepSeek-V4.1-Flash.
For \textbf{RQ3}, we additionally consider MiniMax (V2.7)~\cite{MiniMax}, Qwen-Plus (V3.6)~\cite{Qwen-Plus}, Claude Opus (V4.7)~\cite{Claude-Opus} and GPT (V5.5)~\cite{GPT}.
All models are accessed via the OpenRouter platform~\cite{openrouter}, except 
for DeepSeek, which is accessed via its official API. 
For all models, we use the default settings for maximum token limits and temperature in thinking mode.

\smallskip 
\noindent\textbf{Metrics.}
We report the number of programs for which termination is successfully proved (denoted by \#SP). For each model, we report the average number of input and output tokens (\#inT and \#outT) per benchmark, and their average cost (in USD) calculated based on the official discounted pricing. To evaluate efficiency, we report 
two metrics: the average end-to-end runtime (including LLM latency) over successfully solved programs,
and the penalized average end-to-end runtime (PAR-2) over all programs (lower is better). 
PAR-2 is defined as $\frac{1}{N}\sum_{i=1}^N {\tt T}_i$ for $N$ programs, where ${\tt T}_i$ is the solving time if it is solved or $2\times T$ otherwise with 
$T$ the timeout limit. It is a widely used metric in both hardware verification and SAT competitions for measuring overall performance. 

\smallskip
\noindent\textbf{Budgets.}
For all tools, we impose 
a timeout of 1,800 seconds (wall-clock) per task; as a result, $T=1{,}800$ for PAR-2 calculation.  
For LLM-based tools (i.e., \tool, \toolv, {\sf LOOPY}, {\sf LIFT} and {\sf PROTON}), we impose a 60-second timeout for each safety verification  and a limit of $6$ LLM queries per task. 
Since neither {\sf LOOPY} nor {\sf PROTON} uses any refinement step,
{\sf LOOPY}  (resp. {\sf PROTON}) requires  3--6 (resp.\ 1) LLM queries per solving attempt, 
we allow them up to $2$ and $6$ solving attempts, respectively, with a limit of $6$ LLM queries per task.
In contrast, the other three LLM-based tools employ iterative refinement, thus
are allowed up to five refinement steps following the initial generation.



All experiments were conducted on a machine with an Intel(R) Xeon(R) Platinum 8368Q CPU (76 cores, 2.60 GHz) and running Ubuntu 22.04.5 LTS, with 1024 GiB of RAM available for evaluation.     

 
\subsection{RQ1: Comparison with Baselines}
To answer \textbf{RQ1}, we compare \tool{} against all seven baselines, the fine-tuned LLM {\sf PRO-LLM},
and the variant \toolv, where {\sf LOOPY}, {\sf LIFT}, \tool and \toolv all use the same DeepSeek model.
The results are shown in Table~\ref{tab:compare-tools}, where the best result is in \textbf{bold} and the second best is \underline{underlined}.

\begin{table}[!t]
\caption{Results of comparison over baselines ({\sf RQ 1}).}
\vspace{-1mm}\setlength{\tabcolsep}{2pt}
\centering
\label{tab:compare-tools}
\scalebox{0.89}{\begin{tabular}{l|crr|crr|crr}
\hline
\multirow{2}{*}{\bf Tool} 
& \multicolumn{3}{c|}{\textbf{BS-F} (170)}
& \multicolumn{3}{c|}{\textbf{BS-L} (188)}
& \multicolumn{3}{c}{\textbf{Total} (358)} \\
& \textbf{\#SP} & \textbf{Time(s)} & \textbf{PAR-2}
& \textbf{\#SP} & \textbf{Time(s)} & \textbf{PAR-2}
& \textbf{\#SP} & \textbf{Time(s)} & \textbf{PAR-2} \\ \hline
{\sf AProVE}         & 123 & \underline{23.52} & {1012.31} & 159 & \underline{18.33} & {570.82} & 282 & \underline{20.59} & {780.47} \\
{\sf ddlTerm}        & 140 & 62.95 & {687.13}  & N/A & N/A  & N/A & N/A   & N/A & N/A    \\
{\sf {LIFT}}        & {140} & {128.71} &{741.29} & N/A & N/A  & N/A & N/A   & N/A & N/A    \\
{\sf UAutomizer}     & 116 & \textbf{11.50} &{1151.37}  & {167} & \textbf{5.26}    &  {406.80} & 283 & \textbf{7.82}   & {760.37}  \\
{\sf FreqTerm}       & 80  & 78.66   & {1942.90} & 75 & 28.36    & {2175.14}  & 155 & 54.32    & {2064.86}  \\
{\sf LOOPY}          & {150} & {234.07} & {630.06}  & {164} & {170.43}& {608.24} & {314} & {200.83}   &{618.61} \\
{\sf PROTON}         & 148 & 148.65 &  {595.29}  & 163 & 130.46  & {591.84}& 311 & 139.12   & {593.48}\\
{\sf PRO-LLM}        & 11  & {55.58}   & {3370.65} & 4   & {52.88}   & {3524.53} & 15  & {54.86}   & {3451.46} \\
\hline
\textbf{\toolv}       &  \underline{158}&  60.30 &  \underline{310.17}   & \underline{168} & 36.20 &  \underline{415.33} &  \underline{326} &   47.88 &  \underline{365.39}  \\ 
\textbf{\tool}       & {\textbf{161}} & {66.81} & {\bf 253.86}  & {\textbf{172}} & {46.85} & {\bf 349.24}  &  {\textbf{333}} & {56.50}   & {\bf 303.95}\\ \hline
\end{tabular}}
\begin{justify}\footnotesize 
Note: (1) {\sf ddlTerm} and {\sf LIFT} both require instrumented programs, but their
  artifacts provide such instrumentation only for BS-F. Thus, we do not evaluate
  either tool on BS-L. (2) {\sf FreqTerm} originally supports three CHC solvers 
(FreqHorn, $\mu$Z3 and Spacer). However, due to version compatibility issues between {\sf FreqTerm} and
the solvers, we can only run it with FreqHorn.
\end{justify}
\vspace{-4mm}
\end{table}


\smallskip
\noindent
{\bf Effectiveness.}
Overall, \tool outperforms all seven baselines as well as {\sf PRO-LLM}, solving more tasks on both benchmark suites,
In particular, \tool solves 19 more tasks than {\sf LOOPY}, the best-performing baseline;
and 22 more tasks than {\sf PROTON}, the winner of the SV-COMP 2026 Termination Category.
The fine-tuned small model {\sf PRO-LLM} solves
only 11 tasks on BS-F and 4 on BS-L, all of which are already solvable by {\sf PROTON} without invoking its fine-tuned LLM.
Although the variant \toolv also outperforms all baselines, it solves less programs
than \tool, confirming the benefit of reformulating ranking function synthesis as a code-assertion generation task.

Notably, the authors of {\sf LIFT}~\cite{gong2026llmguided} argued that 
that synthesizing valid ranking functions is
particularly challenging for LLMs because even
minor inaccuracies can invalidate a termination
proof, while LLM-generated expressions are prone
to semantic errors. Consequently, they focus on synthesizing
loop bounds rather than exact ranking functions.
In contrast, our results demonstrate that,
with an appropriately designed framework, LLMs
can effectively synthesize verifiable ranking
functions, which not only establish termination
but also provide a basis for cost analysis
and loop optimization.
 
Table~\ref{tab:aongLLMqueries} shows the number of proved programs 
along with the number of LLM queries; for \textsf{LIFT}, we report only the results on BS-F. 
(We exclude {\sf PRO-LLM} because of its poor effectiveness.)
Both \tool{} and \textsf{LIFT} become stable after five queries (highlighted in {\bf bold}); \toolv{} becomes stable after three queries, while \textsf{LOOPY} becomes stable after six queries (we also tried several other query counts). 
\tool{} already solves more tasks than all baselines after two queries (i.e., one refinement iteration).
The variant \toolv{} performs better than \tool{} within the first two queries, possibly due to benchmark contamination,
but is surpassed by \tool{} after two refinement iterations.  
Compared with \textsf{LIFT}, all 161 programs in BS-F
are proved by \tool within four queries.

\begin{table}[t]
\centering\setlength{\tabcolsep}{6pt}
\caption{Number of proved program along with LLM queries.}\vspace{-1mm}
\label{tab:aongLLMqueries}
\scalebox{0.95}{\begin{tabular}{|l|c|c|c|c|c|c|} \hline
{\bf Tool}  & 1 & 2 & 3 & 4 & 5 & 6 \\ \hline
{\sf LOOPY}  & 0 & 266 & 301 & 309 & 312 & \textbf{314} \\
{\sf LIFT} & 121 & 125 & 128 & 139 & \textbf{140} & 140 \\ \hline
\toolv  & 312 & 323 & \textbf{326} & 326 & 326 & 326 \\
\tool & 302 & 321 & 331 & 332 & \textbf{333} & 333 \\
\hline
\end{tabular}}\vspace{-3mm}
\end{table}

\smallskip
\noindent
{\bf Efficiency.} On the successfully solved programs, \tool is about twice as fast as the other LLM-based baselines (namely, {\sf LOOPY}, {\sf LIFT}, and {\sf PROTON}), but is comparable to or significantly less efficient than the other symbolic or data-driven baselines. Interesting, 
\tool is  only slightly slower than the fine-tuned local model {\sf PRO-LLM}.
Across all programs, \tool achieves the lowest PAR-2 score, confirming
the overall performance of \tool. We later show that \tool{} can achieve higher effectiveness and substantially lower runtime with GPT.
The variant \toolv is slightly more efficient than \toolv on the successfully solved programs and 
achieves a far lower PAR-2 score than all baselines. However, its PAR-2 score is much larger than \tool, e.g., 365.39 vs 303.95 over all benchmarks.

\begin{figure*}[!t]
    \centering
    \includegraphics[width=.95\linewidth]{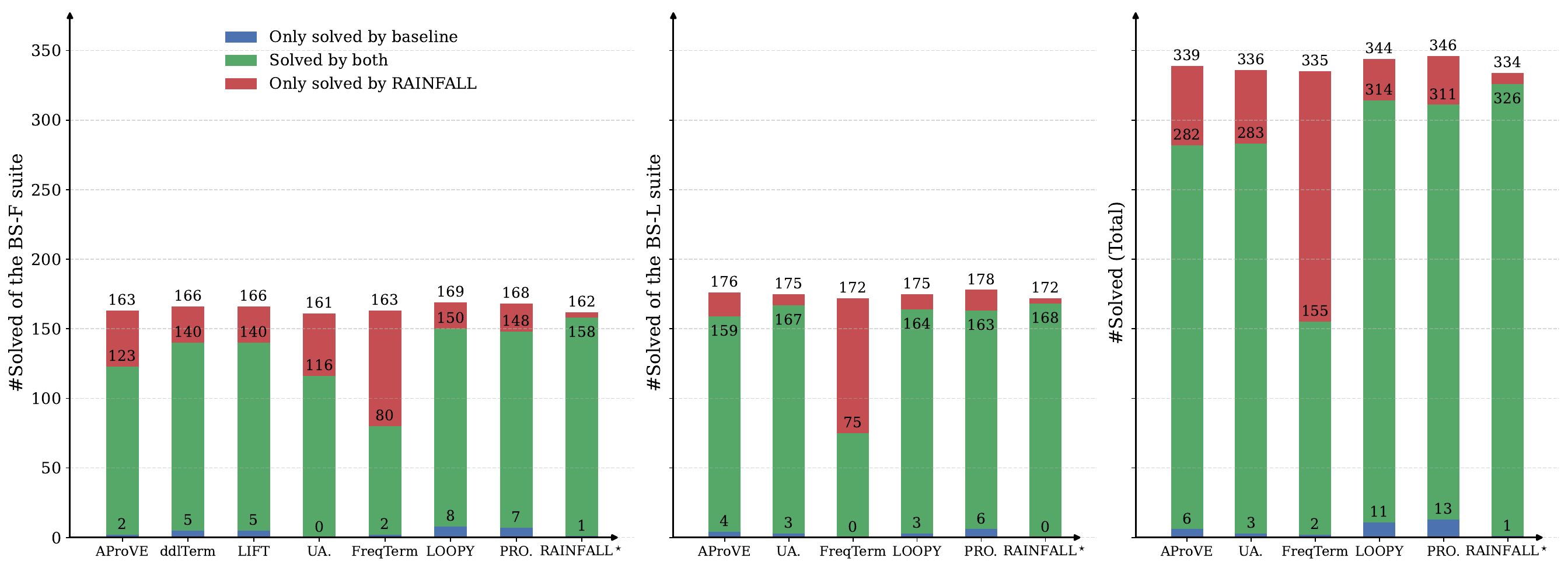}\vspace{-1mm}
    \caption{Distributions of solved tasks, where UA.=UAutomizer and PRO.=PROTON.}
    \label{fig:distribution}\vspace{-4mm}
\end{figure*}

\smallskip
\noindent
{\bf Complementarity.}
Figure~\ref{fig:distribution} depicts the distributions of tasks 
solved exclusively by \tool, exclusively by each baseline, and by both.
We observe that 
\begin{itemize}[left=5pt, topsep=2pt]
    \item the majority of programs are proved by both \tool and the baseline;
    \item \tool uniquely solves much more programs than any baseline or its variant \toolv.
    Moreover, with one exception, every program proved by \toolv{} can also be proved by \tool{}.
    \item  combining \tool{} with the baselines yields the best results.
Particularly, the combination (\tool\hspace{-1mm}+{\sf LOOPY})
  proves {169} (out of 170 programs) on BS-F, for a total of {344} out of 358 programs.
  while the combination (\tool\hspace{-1mm}+{\sf PROTON}) proves {178} (out of 188 programs) on BS-L, for a total of {346} out of 358 programs.
\end{itemize}


\vspace{-1mm}
 \begin{tcolorbox}[notitle,left=0.5mm,right=0.5mm,top=0.2mm,bottom=0.2mm]
 {\bf Answer to RQ1:}
With a single refinement iteration, \tool{} already surpasses all state-of-the-art baselines, including symbolic, data-driven, and LLM-based tools. With four refinement iterations, it stabilizes, solves much more programs, and achieves a significantly lower PAR-2 score. Moreover, 
\tool is complementary to the baselines.
 \end{tcolorbox}\vspace{-3mm}

\subsection{RQ2: Ablation Study}\label{sec:ablations}
To answer {\bf RQ2}, we conduct an ablation study on BS-F to evaluate the contribution of each core component, 
using the full setting described above unless otherwise stated. 


\smallskip
\noindent
{\bf Without few-shot examples.}
We evaluate \tool with and without few-shot examples using only one LLM query
(namely, without refinement) to isolate their impact.
  Without few-shot examples, \tool (i.e., DeepSeek) solves 140 tasks, 5 fewer than the 145 tasks solved with few-shot examples under the same setting,
  demonstrating the effectiveness of few-shot examples in guiding code-assertion generation.


\smallskip
\noindent
{\bf Reject-only feedback.} 
We compare \tool with a variant which feeds back only the rejection result and the incorrect output to the LLM.
This variant solves 155 tasks, 6 fewer than the full \tool, showing
the effectiveness of informative diagnostic feedback for refinement.


\smallskip
\noindent
{\bf Restarting solving attempts instead of refinement.}
We compare \tool with a variant that directly restarts the solving attempt with diagnostic
feedback instead of refining the LLM's incorrect output.
For a fair comparison, this variant can restart at most 6 times.
This variant solve 159 tasks, 2 fewer than \tool.
This small gap indicates that most of the benefit of feedback comes from the diagnostic
feedback rather than from retaining the incorrect output.

\smallskip
\noindent
{\bf Symbolic verification.} 
%
To assess the effectiveness of our symbolic verification,
we count the number of errors reported by the syntactic equivalence, code-assertion completeness and assertion checks when analyzing all 170 programs in BS-F. These checks report 3, 16 and 5 errors in total, respectively,
indicating that they are essential for filtering and diagnosing invalid outputs.
We do not evaluate their contribution to the number of solved programs, because
\tool becomes unsound if any of them is dropped and manually checking the resulting ranking functions is error-prone and time-consuming.

Together with the results reported in Table~\ref{tab:aongLLMqueries}, we answer \textbf{RQ2} as follows.

\vspace{-1mm}
 \begin{tcolorbox}[notitle,left=0.5mm,right=0.5mm,top=0.2mm,bottom=0.2mm]
 {\bf Answer to RQ2:}
Our placeholder contracts, few-shot examples, symbolic verification,
and diagnostic feedback all contribute the performance of \tool. 
 \end{tcolorbox}\vspace{-3mm}

\begin{table}[t]
    \centering
    \caption{Results on BS-F wtih more LLMs ({\bf RQ3}).}
\label{tab:llm-comparison}
\setlength{\tabcolsep}{4pt}
\scalebox{0.95}{ \begin{tabular}{cc|ccccrr}
  \hline
  {\bf Model} & {\bf Size} & {\bf \#SP} & {\bf \#inT} & {\bf \#outT} & {\bf USD} & {\bf Time}(s) & {\bf PAR-2} \\
  \hline
  Claude-Opus &N/A  & 160 & 3600  & \underline{10741} & 0.287 & \underline{93.89}  & 268.18 \\
  DeepSeek  & {552B}  & \underline{161} & \underline{2207}  & {13277} & {\bf 0.015} & {94.90} & \underline{253.86} \\
  GPT       &N/A   & \textbf{163} & \textbf{2065} & \textbf{2668} & {0.090} & \textbf{54.73} & \textbf{187.24} \\
  MiniMax    & 230B  & 156 & {2778}  &{32004} & \underline{0.039} & {457.00} & {735.40} \\
  Qwen-Plus & 397B  & 138 & 7109  & 38132 & 0.077 & 730.53 & 922.35 \\
  \hline 
 \end{tabular}} \vspace{-2mm}
\end{table}

\subsection{RQ3: Impact of LLMs}
To answer {\bf RQ3}, we evaluate \tool on BS-F with five LLMs, including DeepSeek-V4.1-Flash, the default model in {\bf RQ1} and {\bf RQ2}.
The results are presented in Table~\ref{tab:llm-comparison}. Note that model sizes for proprietary LLMs are not publicly disclosed.
Overall, \tool still outperforms all baselines with most LLMs. 
The only exception is \tool\hspace{-1mm}(Qwen-Plus), which performs worse than {\sf ddlTerm}, {\sf LIFT}, {\sf LOOPY} and {\sf PROTON} despite using more tokens and taking longer.
Among the evaluated models, GPT solves the most programs and achieves the lowest average runtime, while DeepSeek and Claude-Opus are close behind. 
Moreover, we observe that  
Claude-Opus incurs substantially higher cost (more than three times that of GPT).  
While GPT is more effective and efficient, we use DeepSeek for {\bf RQ1} and {\bf RQ2}
because of its lower cost. 

Although \tool with Qwen-Plus as the LLM performs worse than {\sf ddlTerm}, {\sf LIFT}, {\sf LOOPY} and {\sf PROTON}, Table~\ref{tab:rq3_complement} shows that \tool consistently improves all baselines across all LLMs, with GPT and DeepSeek achieving the best overall performance.
These results confirm the robustness of \tool across different LLMs.

\begin{table}[t]  
\centering
\caption{Results of baseline+\tool on BS-F.}
\label{tab:rq3_complement}
 \vspace{-2mm}\setlength{\tabcolsep}{4pt} 
\scalebox{0.9}{\begin{tabular}{l|cccccc}
\hline
\textbf{Baseline} & {\bf Sole} & \textbf{Claude-Opus} & \textbf{{DeepSeek}} & \textbf{GPT} & \textbf{MiniMax} & \textbf{Qwen-Plus} \\
  \hline
  {\sf AProVE}     & 123 & 162 & {163} & \textbf{165} & \underline{164} & 159 \\
  {\sf ddlTerm}    & 140 & \underline{164} & {\bf 166} & \underline{164} & 162 & 157 \\
  {\sf LIFT}       & 140 & \underline{164} & {\bf 166} & \underline{164} & 162 & 157 \\
  {\sf UAutomizer} & 116 & \underline{161} & \underline{161} & \textbf{164} & \underline{161} & 155 \\
  {\sf FreqTerm}   & 80  & 162 & \underline{163} & \textbf{164} & 158 & 143 \\
  {\sf LOOPY}      & {150} & {166} & {\bf 169} & {\bf 169} & {166} & {165} \\
  {\sf PROTON}     & {148} & 167 & \underline{168} & \textbf{169} & 167 & 164 \\
\hline
\end{tabular}}\vspace{-2mm}
\end{table}

\subsection{Case Studies}
\noindent 
{\bf Failure cases.}
We manually analyze the 9 cases from BS-F that \tool fails to prove.
  \begin{itemize}[left=5pt, topsep=2pt]
  \item 
  For 5 programs, \tool (with DeepSeek as the default LLM) generates candidate ranking functions for loops involving modular arithmetic or nonlinear multiplicative updates, but fails to verify any of them within the default 60-second timeout. Extending the verification timeout to 600 seconds enables three of these candidates to be verified. One of these 3 programs can also be solved within 60 seconds when \tool uses GPT (instead of DeepSeek). These results confirm that DeepSeek generates valid ranking functions for at least 3 (out of 5) programs, although verifying them requires additional time.
  

      \item 
      

For 3 programs, DeepSeek generates piecewise candidate ranking functions over large integer ranges, which fails to verify within either the default 60-second timeout or an extended 600-second timeout. Nevertheless, when using GPT, \tool successfully proves termination for all three programs within the default 60-second timeout.

  
  \item  
For one program, \tool uses up all five refinement iterations attempting to repair candidates that fail the three checks. The final candidate, which is a $2$-phase ranking function, remains unverified by {\sf UAutomizer} even with an extended timeout of 600 seconds. Switching the underlying LLM to GPT also fails to solve the program within the default 60-second timeout.

\end{itemize}
%
As a result, \tool equipped with DeepSeek and GPT can solve at least 167 out of 170 programs in BS-F within the 600-second verification time limit,
while the remaining three programs can be solved by {\sf PROTON}.

\smallskip
\noindent
{\bf Success cases.}
We analyze the four cases from BS-F on which \tool successfully proves termination but
none of the four conventional baselines (i.e., {\sf AProVE},
{\sf ddlTerm},  {\sf FreqTerm} and {\sf UAutomizer}) succeeds.
  
  \begin{itemize}[left=5pt, topsep=2pt]
      \item {\bf Case 1:} \tool{} generates the valid $4$-lexicographic ranking function $(f_1,f_2,f_3,f_4)$, where
      \[
      \begin{aligned}
        f_1&=\max(z,0), &\quad f_2&=\max(-x,0),\\
        f_3&=x-y, &\quad
        f_4&=\begin{cases}
          1+\max(z,-2)+2, & x=0,\\
          \max(z,-2)+2, & x\neq0.
        \end{cases}
      \end{aligned}
      \]
      {\sf AProVE} and {\sf ddlTerm} both time out, {\sf FreqTerm} returns \emph{no candidate}, and {\sf UAutomizer} returns {\it unknown}.
     
     %
      \item {\bf Case 2:} \tool{} generates the
      $2$-phase ranking function $(f_1',f_2')$, where
      \[
      \begin{aligned}
        f_1'&=n-|z|, \quad 
        f_2'=\begin{cases}
          2, & A\wedge z>0\wedge B,\\
          1, & A\wedge\neg(z>0\wedge B),\\
          0, & \neg A,
        \end{cases} \\
        A&\coloneqq|z|>n\wedge x+y\ge0\wedge x\le n \\
        B&\coloneqq2x-y+z\ge0\wedge 2x-y\le n\\
      \end{aligned}
      \]
      {\sf AProVE} returns \textit{maybe}, {\sf ddlTerm} times out, and {\sf FreqTerm} returns \emph{no candidate}, and
       {\sf UAutomizer} returns {\it unknown}.
    %
      \item {\bf Case 3:} \tool{} generates the $3$-phase
      ranking function $(f_1,f_2,f_3)$, where
      \[
        f_1=z+1,\qquad f_2=y,\qquad f_3=x+1.
      \]
 {\sf AProVE} and {\sf ddlTerm} both time out, {\sf FreqTerm} returns \emph{no candidate}, whereas
{\sf UAutomizer} throws an \textit{AssertionError}.
      \item {\bf Case 4:} \tool{} generates the single affine ranking
      function $f=N-x+4-4z$.
      %
{\sf AProVE} and {\sf ddlTerm} both time out as well, {\sf FreqTerm} returns \emph{no candidate}, and {\sf UAutomizer} runs out of memory.
  \end{itemize}

These cases highlight the flexibility of \tool in generating a diverse range of ranking functions without being constrained by predefined templates that fix either the number of components or the grammar of ranking function expressions.
In contrast, the failures of conventional tools may be attributed to the difficulty of identifying suitable ranking functions within the allocated resource budgets or to the restrictions of their predefined templates.

%
%
We further examine the five BS-F cases that \tool{} solves but neither of the two LLM-based baselines, {\sf LIFT} and {\sf LOOPY}, solves.
For these cases, \tool{} successfully generates two 2-phase ranking functions, two 3-lexicographic ranking functions, and one 2-lexicographic ranking function. 
{\sf LIFT} prompts only for conjunctive and lexicographic loop bounds,
but either fails to find a verified solution within the given budget (3600 seconds in total, 60 seconds per safety verification, and 6 LLM queries) or encounters runtime errors.
Meanwhile, {\sf LOOPY} generates lexicographic and multiphase
candidate ranking functions, but Frama-C rejects all of them
within the same budget.
Conversely, {\sf LIFT} and {\sf LOOPY} together solve eight BS-F tasks that
  \tool{} does not, all of which belong to the nine failure cases analyzed above.
These results highlight the complementary strengths of \tool and existing LLM-based approaches.

\subsection{Threats to validity}
Our approach relies on both third-party safety verifiers and LLMs. We use {\sf UAutomizer}~\cite{HeizmannHP14} as the safety verifier because it has been actively maintained for over a decade and ranked first overall in SV-COMP from 2023 to 2026.
Although other safety verifiers could be integrated into our framework, the performance may vary depending on the verifier employed.

To mitigate the impact of LLMs, we evaluate \tool using a diverse set of both open-source and proprietary models. \tool outperforms all baselines with most LLMs; the only exception is Qwen-Plus, for which \tool performs worse than {\sf ddlTerm}, {\sf LIFT}, {\sf LOOPY} and {\sf PROTON}. Nevertheless, \tool consistently improves upon all baselines across all evaluated LLMs.

We evaluate \tool on two widely used public benchmark suites, and its effectiveness may vary on other benchmarks or real-world programs. 
Since modern LLMs are typically trained on large corpora containing publicly available code,
these benchmarks with ranking functions may have been included in their training data, raising the possibility of benchmark contamination. 
Nevertheless, we reformulate the ranking function synthesis as a code-assertion generation task,
and, to the best of our knowledge, no public benchmark contains the code-assertions used by our framework.
Moreover, as shown in Section~\ref{sec:ablations}, DeepSeek alone solves only 140 out of 170 tasks in BS-F, comparable to state-of-the-art baselines,
whereas our approach raises this number to 161. Evaluating \tool{} on additional benchmark suites and real-world programs remains an important direction for future work.

\section{Related Work} \label{sec:related}
 
\noindent{\bf Single ranking function synthesis.}
The most standard approach to proving program termination is via ranking functions that decrease along every loop iteration.
Podelski and Rybalchenko~\cite{podelski2004complete} gave a complete method for synthesizing linear ranking functions on linear loops.
Later work relaxes the strict decrease requirement, e.g., by eventual linear ranking functions~\cite{BagnaraM13}.
These approaches typically reduce the synthesis problem to constraint solving (via, e.g., Farkas' lemma), but are limited in expressiveness.

\smallskip
\noindent
{\bf Template-based ranking function synthesis.}
Template-based approaches synthesize ranking functions by reducing the problem to constraint solving over predefined templates.
For a chosen ranking function class, a parameterized template with unknown coefficients is fixed, and synthesis amounts to finding coefficient values that satisfy the required conditions.
Common templates include linear~\cite{LeikeH15}, multiphase~\cite{BenAmramG17}, and lexicographic ranking functions~\cite{BradleyMS05,icalp/BradleyMS05,AliasDFG10,Ben-AmramG14}, as introduced in Section~\ref{sec:preliminaries}.
For example, to synthesize a single linear ranking function for a loop with guard $\phi$, one can use an affine template $f(\vec{x}) = \vec{a}\cdot\vec{x} + c$, where $\vec{a}$ and $c$ are unknown.
The synthesis task reduces to determining $\vec{a}$ and $c$ such that the ranking function conditions hold for all program transitions and $f(\vec{x})<0 \implies \neg \phi$.
Leike and Heizmann~\cite{LeikeH15} reduce the synthesis problem of common ranking functions above to solving nonlinear constraints via SMT.
More expressive templates may use polynomial or rational expressions, or multiple components as in multiphase and lexicographic ranking functions. Thus, the effectiveness of template-based methods depends heavily on choosing suitable templates, a task that aligns well with the strengths of LLMs.

\smallskip
\noindent
{\bf Data-driven techniques.}
Some learning-based approaches predict termination without formal guarantees~\cite{AlonD22}, while others adopt a learn-and-check paradigm~\cite{Nori013,UrbanGK16,xu2022data,LiLLSTZ22}.
Neural approaches, in particular, learn candidate functions from execution traces and validate them using SMT solving to ensure soundness~\cite{GiacobbeKP22,AbateGR20}. These techniques heavily rely on the quality and quantity of execution traces for training neural networks.
\cite{GiacobbeKPT25} trains neural networks to simultaneously represent ranking functions and inductive invariants. 
Unlike earlier approaches, they do not require execution traces. Instead, the neural networks are trained through a sequence of 
NP-complete MILP solving in a counterexample-guided inductive synthesis loop and are designed for hardware verification 
rather than proving termination of software programs.  

\smallskip
\noindent
{\bf Other symbolic approaches.}
Alternative approaches reason about transition relations and exploit transition invariants.
Key techniques include size-change analysis~\cite{DBLP:conf/popl/LeeJB01}, refinement~\cite{ChawdharyCGSY08,CookPR06}, variance analysis~\cite{BerdineCCDO07,PodelskiR05,KroeningSTW10}, abstract interpretation~\cite{CousotC12}, temporal reasoning~\cite{LeGHC14,LeQC15}, and interprocedural summarization~\cite{ChenDKSW18}.
Some techniques further address complex data structures, e.g., heaps~\cite{BerdineCDO06}, pointers~\cite{CookPR06}, and lists~\cite{HenselG23}.

\smallskip
\noindent
{\bf Termination checking tools.}
{\sf AProVE}~\cite{LommenG25} uses integer transition systems (ITS) and integrates \textsc{KoAT}~\cite{DBLP:journals/corr/abs-2512-18851} for inferring complexity bounds of ITS.
Both {\sf UAutomizer}~\cite{HeizmannHP14,ChenHLLTTZ18} and \textsf{T2} (and its predecessor {\sf Terminator})~\cite{CookPR06,CookSZ13,BrockschmidtCIK16} iteratively prove termination by analyzing cycles; {\sf UAutomizer} employs automata-based reasoning to discover new cycles, while \textsf{T2} reduces the problem to safety checking.
{\sf FreqTerm}~\cite{FedyukovichZG18} enumerates candidate ranking functions.
{\sf PROTON 2.1}~\cite{mukhopadhyay2025proton} leverages fine-tuned LLMs to handle cases not solved by \textsf{PROTON 1.1.2}~\cite{MettaKMVC24}, a bounded model checking-based tool. 
{\sf LOOPY}~\cite{kamath2024leveraging} prompts LLMs to generate loop invariants and candidate ranking functions directly from natural-language definitions.
When a candidate is incorrect, {\sf LOOPY} initiates a new interaction with the LLM without incorporating the verification feedback.
%
{\sf ddlTerm}~\cite{xu2022data} adopts a data-driven approach to learn loop bounds.
The concurrent work {\sf LIFT}~\cite{gong2026llmguided}
presents a guess-and-check workflow to learn loop bounds.
Both {\sf ddlTerm} and {\sf LIFT} require programs to be manually instrumented for each type of loop bounds
and moreover, loop bounds only provide an upper bound on the maximum number of iterations a loop may execute.
They are therefore not exact. In contrast, ranking functions are exact and amenable to cost analysis and loop optimization. 
Our approach properly addresses this challenge and achieves better results than {\sf ddlTerm} and {\sf LIFT} without requiring manual instrumentation.
Other tools include {\sf DynamiTe}~\cite{LeAFKN20} and {\sf FuncTion}~\cite{Urban15}.
Related work also addresses non-termination~\cite{GuptaHMRX08,ChenCFNO14,HanH23} and LLM-based prediction~\cite{SAKVSAW26}, which is not our focus.
\section{Conclusion} \label{sec:conc}

We present \tool, a novel neuro-symbolic framework for ranking function synthesis. We formulate ranking function synthesis as a constrained code-assertion generation task with dedicated placeholders, integrating LLM-based generation with symbolic verification through a generate–verify–refine paradigm. We show that incorporating ranking-function expertise, few-shot examples, and verification diagnostics into LLM prompts significantly improves synthesis effectiveness. To ensure soundness and provide fine-grained diagnostics, we develop a symbolic verification pipeline that validates LLM-generated code and assertions and provides actionable feedback for iterative refinement. Experimental results demonstrate that \tool outperforms state-of-the-art baselines and achieves further improvements when combined with existing symbolic termination provers. These findings highlight the potential of tightly integrating LLM-based generation with symbolic reasoning for automated termination proving.




\section*{Acknowledgment}
This work was supported by the Strategic Priority Research Program of the Chinese
Academy of Sciences, Grant No. XDA0320101, and partially supported by NSFCRGC Collaborative Research Grant No. 62561160151. T. Chen is partially supported by overseas grants from the State Key Laboratory of Novel Software Technology, Nanjing University (KFKT2025A05, KFKT2026A04)

\balance
\bibliographystyle{IEEEtran}
\bibliography{acmart}

\end{document}